\documentclass{article}
\usepackage{spconf,amsmath,graphicx,amssymb}
\usepackage{subcaption}
\usepackage{multirow}
\usepackage{booktabs}
\usepackage[colorlinks=false,pdfborder={0 0 0}]{hyperref}
\usepackage[T1]{fontenc}

\usepackage{mathtools}

\usepackage{xfrac}

\newcommand{\D}[1]{\mathrm{d}{#1}}
\newcommand{\bX}{\mathbf X}

\newcommand{\bY}{\mathbf Y}

\usepackage{comment}

\title{EMOCONV-DIFF: Diffusion-based Speech Emotion Conversion for Non-parallel and In-the-wild Data}
\name{Author(s) Name(s)\thanks{Thanks to XYZ agency for funding.}}
\address{Author Affiliation(s)}

\name{Navin Raj Prabhu$^{\star \dagger}$ \quad Bunlong Lay$^{\star}$ \quad Simon Welker$^{\star}$ \quad Nale Lehmann-Willenbrock$^{\dagger}$  \quad Timo Gerkmann$^{\star}$\thanks{This work was funded under the Excellence Strategy of the Federal Government and the Länder, and the ''Mechanisms of Change in Dynamic Social Interaction'' project (LFF-FV79, Landesforschungsförderung Hamburg).}}

\address{$^{\star}$Signal Processing, Universit\"at Hamburg, Germany \\
      $^{\dagger}$Industrial and Organizational Psychology, Universit\"at Hamburg, Germany \\
          \texttt{navin.raj.prabhu@uni-hamburg.de}}

\begin{document}
\topmargin=0mm
\ninept
\maketitle
\begin{abstract}
Speech emotion conversion is the task of converting the expressed emotion of a spoken utterance to a target emotion while preserving the lexical content and speaker identity. While most existing works in speech emotion conversion rely on acted-out datasets and parallel data samples, in this work we specifically focus on more challenging in-the-wild scenarios and do not rely on parallel data. To this end, we propose a diffusion-based generative model for speech emotion conversion, the EmoConv-Diff, that is trained to reconstruct an input utterance while also conditioning on its emotion. Subsequently, at inference, a target emotion embedding is employed to convert the emotion of the input utterance to the given target emotion. As opposed to performing emotion conversion on categorical representations, we use a continuous arousal dimension to represent emotions while also achieving intensity control. We validate the proposed methodology on a large in-the-wild dataset, the MSP-Podcast v1.10. Our results show that the proposed diffusion model is indeed capable of synthesizing speech with a controllable target emotion. Crucially, the proposed approach shows improved performance along the extreme values of arousal and thereby addresses a common challenge in the speech emotion conversion literature.
\end{abstract}


%
\begin{keywords}
Speech emotion conversion, diffusion models, non-parallel samples, arousal, in-the-wild
\end{keywords}

\section{Introduction}
\label{sec:intro}

Speech is one of the key social signals used by humans to express their emotions \cite{triantafyllopoulos2023overview}. While significant developments have been made in speech generation and synthesis, \textit{emotion-conditioned} speech synthesis is still a challenge \cite{triantafyllopoulos2023overview, amiriparian2023guest}. In the context of human-machine interaction, to improve the naturalness of machine communication, the generation of emotionally expressive speech is required \cite{triantafyllopoulos2023overview}. Speech emotion conversion (SEC) is a sub-field of emotion-conditioned speech synthesis that aims to map a speech signal into another speech signal by converting its emotional expression while preserving the lexical information and the speaker's identity \cite{VCdu22c_interspeech}. 


Emotions are represented in SEC as either \textit{categorical} (e.g., six basic emotions \cite{ekman1971constants}) \cite{VCdu22c_interspeech, zhou2023_emointensitycontrol} or \textit{continuous} (e.g., circumplex model \cite{russell1980circumplex}) \cite{hifigan_rajprabhu23} representations. It is well established in the speech emotion recognition (SER) and psychology literature that emotion is a complex construct with \textit{fuzzy} class boundaries \cite{lotfian2017building}, and the categorical representations (e.g., happy, anger) do not aptly capture the subtle difference between human emotions \cite{russell1980circumplex}. The circumplex model contrarily represents emotions using \textit{continuous} and independent dimensions, i.e., \emph{arousal} (relaxed vs. activated) and \emph{valence} (positive vs. negative) \cite{russell1980circumplex}. While the audio modality typically captures the arousal dimension of emotion well, it insufficiently explains valence \cite{deoliveira2023leveraging, prabhu22_interspeech}. Therefore, in this work, we follow \cite{hifigan_rajprabhu23} and represent emotion using the continuous arousal dimension. Moreover, by using the continuous representations (arousal on a scale of 1 to 7) we directly achieve intensity control in SEC, as opposed to an additional effort required for categorical representations (e.g., \cite{zhou2023_emointensitycontrol, tang23_emomix}).



Current SEC systems are typically trained on high-quality recorded speech data that are \textit{acted-out} by professional actors. As a consequence the resulting algorithms are typically sensitive to noise and variabilities pertinent in real-world scenarios \cite{kzhou23_thesis} (e.g., acoustic noise, speaker variabilities, subtle intonations, or vocal bursts that carry emotion; e.g., \cite{busquet2023voice}). Furthermore, SEC systems trained on acted-out speech may create stereotypical portrayals of emotions \cite{kzhou23_thesis}. Another crucial drawback of acted-out datasets is that they require \textit{parallel} utterances, i.e., each source utterance is required to also have a ground-truth utterance of a target emotion \cite{rizos2020stargan, zhour2020CycleGAN}. However, parallel utterances are expensive to collect \cite{rizos2020stargan}, and models trained on them lack scalability \cite{triantafyllopoulos2023overview}. In this work, we address these drawbacks of acted-out and parallel data by specifically focusing on non-parallel \textit{in-the-wild} data.


A challenge in overcoming the usage of parallel utterances is the problem of \textit{disentanglement}, where a disentanglement technique is required to decompose the source utterance into several constituents (i.e., emotion, lexical, and speaker information) before synthesizing speech for a target emotion \cite{triantafyllopoulos2023overview, VCdu22c_interspeech}. Existing works have employed encoder-decoders \cite{zhou2023_emointensitycontrol}, generative adversarial networks \cite{rizos2020stargan}, and self-supervised learning (SSL) \cite{hifigan_rajprabhu23} for the disentanglement. Recently, the so-called \emph{diffusion models} have been introduced for the synthesis of high-quality samples, both in the audio- and image-domain \cite{richter23_setasl, Wu_2023_CVPR}. Further in \cite{Wu_2023_CVPR}, the disentanglement capability of diffusion models was uncovered for the task of text-conditioned image editing and demonstrated strong control over the image synthesis process. 



\begin{figure*}[t!]
	\centerline{\includegraphics[width=.8\linewidth]{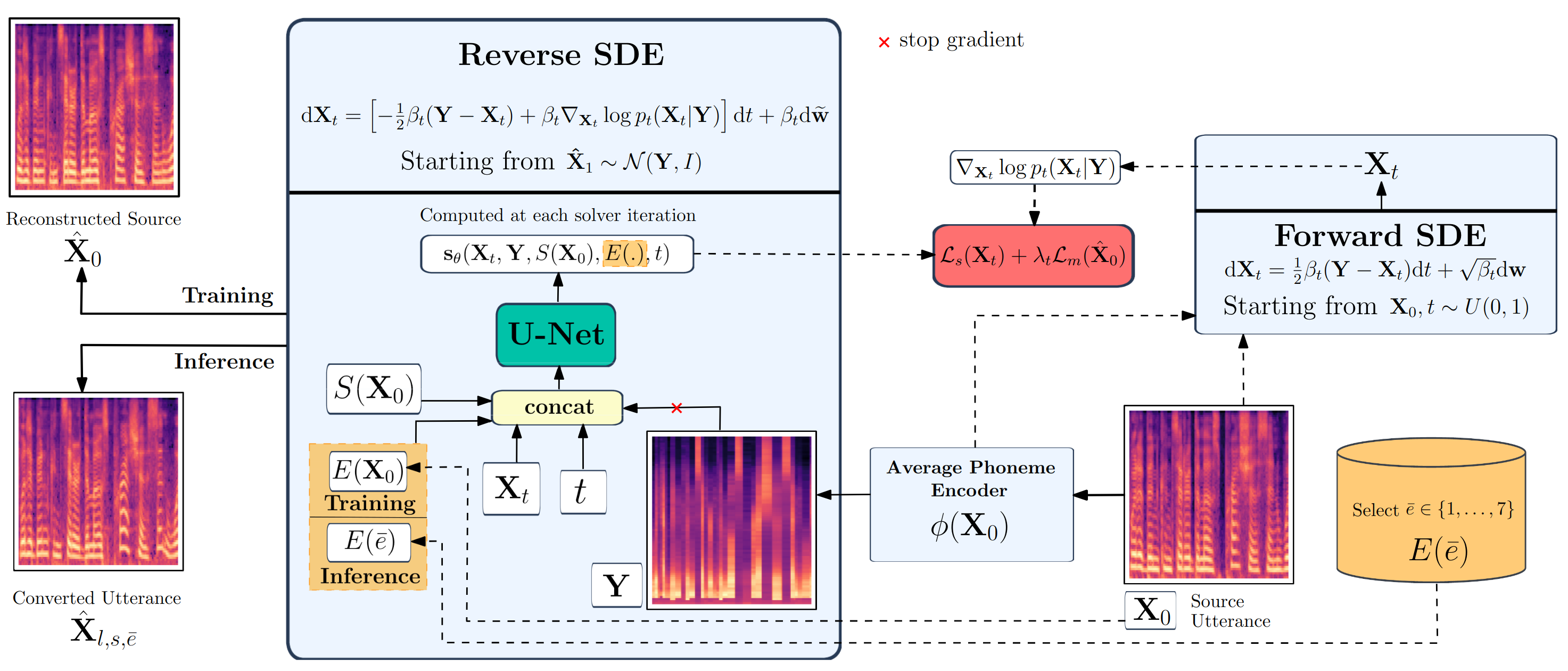}}
	\caption{Illustration of the training and inference process of the proposed EmoConv-Diff approach. Dotted arrows denote operations performed only during training. The \textit{stop gradient} function stops the accumulation of the gradients of the inputs during the training.}
	\label{fig:emovc_architecture}%
\end{figure*} 


For in-the-wild SEC without relying on parallel utterances, we introduce a diffusion-based approach that is trained to reconstruct a source utterance while also conditioning on its emotion. Subsequently, at inference, a target emotion embedding is employed to convert the emotion of the source utterance to the given target emotion. As such, the contributions of this paper are as follows: We introduce a novel emotion-conditioned diffusion model that does not rely on parallel utterances for SEC, which is in contrast to existing emotion-conditioned diffusion models that rely on parallel utterances and operate on the text-to-speech (TTS) domain \cite{guo2023emodiff, tang23_emomix}. Building up on our previous work \cite{hifigan_rajprabhu23}, our models can cope with unseen real-world scenarios, as it is trained on non-parallel in-the-wild speech utterances. To the best of our knowledge, we are the first to tackle this problem of non-parallel in-the-wild data for SEC, and the paper at hand is the first to employ diffusion models for this. Finally, the proposed approach improves over the \textit{HiFiGAN} \cite{hifigan_rajprabhu23} for extreme target emotions, a common problem in SEC and TTS \cite{hifigan_rajprabhu23, wang23ka_interspeech}.


\section{Diffusion Models}
\label{sec:background}

\emph{Diffusion models} are used in various applications across domains for the task of generation, such as image editing \cite{Wu_2023_CVPR}, speech enhancement \cite{lay202interspeech}, and TTS \cite{popov2021grad}. The idea behind these models involves adding Gaussian noise to the data using a stochastic differential equation (SDE). The \emph{forward SDE} or \emph{forward process} can be viewed as transforming an initial distribution into a terminating distribution that is usually tractable and available during inference. Under mild constraints, a forward SDE can be inverted by the \emph{reverse SDE} \cite{anderson1982reverse}. The reverse SDE or \emph{reverse process} transforms the terminating distribution of the forward process back into the initial distribution, during which the disentanglement is achieved \cite{Wu_2023_CVPR}. 

In the extant literature, emotion-conditioned diffusion models rely on parallel data and operate on the TTS domain \cite{tang23_emomix, guo2023emodiff}. In \cite{guo2023emodiff}, the GradTTS-based \textit{EmoDiff} was introduced. EmoDiff achieves emotion-conditioned speech synthesis from source text using a soft-label guidance technique in the reverse process. \cite{tang23_emomix} introduces \textit{EmoMix}, which uses pretrained SER embeddings of a reference utterance to exemplify the target emotional prosody and condition on the desired emotion. Note that both \cite{guo2023emodiff} and \cite{tang23_emomix} rely on acted-out parallel utterances and operate on the TTS domain. 

\section{Proposed Methodology: EMOCONV-DIFF}
\label{sec:methodology}



We define the SEC task as follows: given the mel spectrogram of a source speech utterance $\mathbf{X}_{l, s, e}$ (or simply $\bX_0$), containing lexical content $l$, speaker identity $s$, and emotion information $e$, we aim to generate a new mel spectrogram $\hat{\mathbf{X}}_{l, s, \Bar{e}}$ that only transforms the arousal information to a target value $\Bar{e}$. For this, we introduce a diffusion-based approach, the \textit{EmoConv-Diff}, which is summarized in Fig. \ref{fig:emovc_architecture}. The EmoConv-Diff comprises a set of \textit{encoders}, each encoding the attributes to be disentangled, and a diffusion-based \textit{decoder}, which aims to disentangle the attributes and perform emotion-controllable speech synthesis. The output of the diffusion decoder is a mel spectrogram $\hat{\mathbf{X}}_{l, s, \Bar{e}} \in\mathbb{R}^{n\times T}$ and it is converted into time domain speech signal using a pretrained HiFiGAN vocoder \cite{hifigan}.

\subsection{Encoders}
The EmoConv-Diff comprises three encoders: the \textit{phoneme encoder} $\phi(.)$, 
the \textit{speaker encoder} $S(.)$, 
and the \textit{emotion encoder} $E(.)$.

\noindent\textbf{Phoneme Encoding:} Speaker- and emotion-independent "average voice" phoneme-level mel features are used to encode the lexical content $l$. Let $\bY \coloneqq \phi(\mathbf{X}_0)$ be the "average voice" representation of the source audio, where $\phi(.)$ is the pretrained phoneme encoder. The transformer-based encoder, adopted from \cite{diffvc_popov}, has been used previously in voice conversion tasks. The encoder output (see $\bY$ in Fig. \ref{fig:emovc_architecture}) has the same dimensions as the source mel $\mathbf{X}_0 \in\mathbb{R}^{n\times T}$.

\noindent\textbf{Speaker Encoding:} To encode the speaker identity, we use a pretrained speaker verification model $S(.)$ \cite{jia2018transfer}, following \cite{diffvc_popov}. The output of $S(.)$ is a $d-vector$ speaker representation $S(.) \in \mathbb{R}^{128}$.

\noindent\textbf{Emotion Encoding:} 
To encode emotional information, we use a pretrained SSL-based SER system $E(.) \in \mathbb{R}^{1024}$, introduced in \cite{wagner2023dawn}. The $E(.)$ network was built by fine-tuning the Wav2Vec2-Large-Robust network \cite{wagner2023dawn} on the MSP-Podcast (v1.7) dataset \cite{lotfian2017building}.

\subsection{Diffusion-based decoder}
The diffusion-based decoder follows the SDE formalism by \cite{popov2021grad}. Specifically, let $t$ be the continuous diffusion time-step variable describing the progress of the diffusion process. For $0 \leq t \leq 1$ the forward SDE of this work is given by

\begin{equation} \label{eq:fsde}
    \D{\mathbf \bX_t} =
       \frac{1}{2} \beta_t ( \bY - \mathbf X_t )\D{t}
        + \sqrt{\beta_t} \D{{\mathbf w}},
\end{equation}
where $\mathbf w$ is the standard Wiener process \cite{kara_and_shreve}, $\mathbf \bX_t$ is the current process state with initial condition $\mathbf \bX_0 = \mathbf{X}_{l, s, e}$ and $\beta_t$ is a non-negative function called the noise schedule.
The process state $\mathbf \bX_t$ follows a Gaussian distribution \cite[Section 5]{kara_and_shreve} that is called the \emph{perturbation kernel}:
\begin{equation}
\label{eq:perturbation-kernel}
    p_{0t}(\mathbf \bX_t|\mathbf \bX_0, \mathbf \bY) = \mathcal{N}_\mathbb{C}\left(\mathbf \bX_t; \boldsymbol \mu(\mathbf \bX_0, \mathbf \bY, t), \sigma(t)^2 \mathbf{I}\right).
\end{equation}
The mean evolution of $\mu(\mathbf \bX_0, \mathbf \bY, t)$, or simply $\mu(t)$, is given by
\begin{equation} \label{eq:mean}
    \mu(\mathbf \bX_0, \mathbf \bY, t) = \alpha_t\bX_0 + \left(1- \alpha_t\right) \bY,
\end{equation}
where $\alpha_t = e^{-\frac{1}{2} \int_{0}^t \beta_sds}$ and the variance evolution is given by
\begin{equation}
    \sigma(t)^2 = \left(1 - \alpha_t^2 \right) \mathbf{I}
\end{equation}
We represent the closed-form of $\alpha_t$ as $\beta_t$ and set $\beta_t = b_0 + t(b_1 - b_0)$ and chose $b_0, b_1 > 0$ such that $ \alpha_1 \approx 0$. In this case, the mean evolution describes an interpolation starting at $t=0$ at the distribution of source $\bX_0$ and terminating approximately at the distribution of "average voice" phoneme features $\bY$ at $t=1$. The forward SDE \eqref{eq:fsde} has an associated reverse SDE \cite{anderson1982reverse}:
\begin{equation}\label{eq:plug-in-reverse-sde}
    \D{\mathbf \bX_t} =
        \left[
            - \frac{1}{2} \beta_t ( \bY - \mathbf X_t ) + \beta_t \mathbf  \nabla_{\mathbf \bX_t} \log p_t(\mathbf \bX_t|\mathbf \bY)
        \right] \D{t}
        + \beta_t \D{\widetilde{\mathbf w}}\,,
\end{equation}
where $\D{\widetilde{\mathbf w}}$ is a Wiener process going backward through the diffusion time-steps. Moreover, the reverse process follows the same trajectory as the forward process, i.e. the reverse SDE starts approximately with the distribution of average-voice and terminates for $t=0$ into the distribution of source-targets.

A network called the \emph{score model} $\mathbf s_\theta(\mathbf \bX_t, \mathbf \bY, S(\bX_0), E(\bX_0), t)$, or simply $\mathbf s_\theta(\mathbf \bX_t, t)$, is \textit{trained} to approximate the \emph{score function} $\nabla_{\mathbf X_t} \log p_t(\mathbf \bX_t|\mathbf \bY)$, i.e., the gradients of log-density of noisy data $\bX_t$. We use the U-Net architecture from \cite{diffvc_popov} as the score model $\mathbf s_\theta$. With the trained $\mathbf s_\theta$, we can then use the reverse SDE to generate an estimate of the source target $\mathbf{X}_0$ from the "average voice" $\mathbf \bY$ given speaker identity $S(\bX_0)$ and emotion embeddings $E(\bX_0)$. An intuition behind the reverse process is that the diffusion-based decoder is trained to reconstruct $\bX_0$ while learning the disentanglement between the speech attributes $l$, $s$, and $e$. With this setup, we overcome the need for parallel data during the training process. 

During \textit{inference}, a target emotion embedding $E(\Bar{e})$ is employed to convert the emotion of the source utterance to the given target emotion. The target emotion embedding $E(\Bar{e})$ is defined as the \textit{averaged} emotion embedding of a set reference utterance samples belonging to the emotion category $\Bar{e}$, as

\begin{equation} \label{eq:Etarget}
    E(\Bar{e}) \coloneqq \frac{1}{|A_p(\Bar{e})|}\sum_{\bX_0 \in A_p(\Bar{e})} E(\bX_0),
\end{equation}
where the set of reference samples $A_p(\Bar{e})$ is defined to be the top $p=20\%$ samples belonging to the particular target arousal $\Bar{e}$.

\subsection{Loss functions}
The score model is trained on the \textit{score matching} loss \cite{song20score} which aims to approximate the score function. The score matching loss for $\bX_0$ at time $t$ is formulated as
\begin{equation}
    \mathcal{L}_{s}(\bX_t) = \mathbb{E}_{\epsilon_t} \left[|| \mathbf s_\theta(\mathbf \bX_t, t) + \sigma(t)^{-1} \epsilon_t ||_2^2\right]
\end{equation}
where $\bX_t = \mu(t) + \sigma(t)\epsilon_t$ and $\epsilon_t$ is sampled from $\mathcal{N}(0, \sigma(t))$. In addition to $\mathcal{L}_{s}$, we follow \cite{hifigan_rajprabhu23, tang23_emomix} to use a mel spectrogram reconstruction loss for better conditioning on emotion attributes. $\mathcal{L}_{m}$ measures the $L_1$-norm: 
\begin{align}
    &\mathcal{L}_{m}(\hat{\bX}_0) = \sum_{x} ||\bX_0 - \hat{\bX}_0||_1,
\end{align}
where $\hat{\bX}_0$ is the mel spectrogram of synthesized speech. Note here that during the training of the score model it is expensive to obtain $\hat{\bX}_0$ which requires solving the full reverse SDE. For this, in contrast to \cite{tang23_emomix}, we utilize a single-step approximation of $\hat{\bX}_0$ by only relying on $\bX_t$, $\mathbf s_\theta$, and $\bY$, which are available during training. We use Tweedie's formula \cite{efron2011tweedie} to approximate $\hat{\bX}_0$ as
\begin{equation} \label{eq:X0}
    \hat{\bX}_0 = \dfrac{\hat{\mu}(t) - \left(1-\alpha_t \right) \bY}{\alpha_t} \,.
\end{equation}
where $\hat{\mu}(t)$ is an estimate of $\mu(t)$ \eqref{eq:mean}, and is formulated as $\hat{\mu}(t) = \bX_t - (\mathbf s_\theta(\mathbf \bX_t, t) * \sigma(t)^2)$. With that, the final loss function is
\begin{equation}
    \mathcal{L}(\bX_t, \hat{\bX}_0) = \mathcal{L}_{s}(\bX_t) + \lambda_t \mathcal{L}_{m}(\hat{\bX}_0),
\end{equation}
where $\lambda_t$ is a weighting function depending on the current diffusion time-step $t$. Considering that $\bX_t$ contains more Gaussian noise for larger $t$, we set $\lambda_t = 1 - t^2$, thereby weighting more for smaller $t$ values and gradually decreasing the weights for larger $t$.


\section{Experimental Setup}
\label{sec:experimental-setup}

\begin{table}[t!]
    \centering
        \begin{tabular}{lcl|cc}
    \hline
    \multirow{2}{*}{} & \multicolumn{2}{c|}{DNSMOS $\uparrow$} & \multicolumn{2}{c}{SER Error $\downarrow$}       \\
                                      & SIG             & OVRL         & $\mathcal{L}_{mse}$   &   $\mathcal{L}_{abs}$      \\ \midrule
        
        HiFiGAN \cite{hifigan_rajprabhu23} & \textbf{3.21}  & \textbf{2.79}  & {0.084} &   {24$\%$} \\ 

        $\mathcal{L}_{s}$ & {3.20}  & {2.78}  & {0.091} &   {25$\%$} \\ 

        $\mathcal{L}_{s} + \mathcal{L}_{m}(\bX_t)$ & {3.08}  & {2.62}  & {0.121} &   {34$\%$}\\ 
        
        $\mathcal{L}_{s} + \mathcal{L}_{m}(\bX_0)$ & \textbf{3.21}  & {2.78} & \textbf{0.072}$^*$ &   \textbf{21$\%$}$^*$ \\

        \bottomrule
        \end{tabular}
    \caption{Overall performance of model versions. * indicates statistically significant improvements in results.}
    \label{tab:overall_compare}
\end{table}

\noindent\textbf{Dataset:} The proposed methodology is trained and validated on the \emph{in-the-wild} MSP-Podcast dataset (v1.10) \cite{lotfian2017building}. The dataset in contrast to predominant SEC datasets (e.g., ESD \cite{zhou2022emotionalESD}, IEMOCAP \cite{busso2008iemocap}) is larger ($\approx$238hrs of audio), has utterances of variable duration, has over 1400 speakers, and contains naturalistic emotional expressions. For example, the ESD contains acted-out utterances from only 10 English speakers and only $\approx$29 hours of acted-out utterances. The arousal annotations, collected at the utterance-level on a scale of 1 to 7, are distributed with $\mu=$4 and $\sigma=$0.95.




\noindent\textbf{Validation measures:} We validate the proposed methodology in terms of both the SEC capabilities and the speech quality of the synthesized signal. As the measure of SEC capability, we use the mean-squared ${L}_{mse}$ and mean-absolute ${L}_{abs}$ errors, calculated between the target arousal $\Bar{e}$ and the SER prediction on the synthesized output $E(\hat{\bX})$. As the measure of speech quality, we use the DNSMOS \cite{reddy2022dnsmos}, a non-intrusive objective speech quality metric designed to predict the mean-opinion score (on a scale of 1 to 5) results of subjective listening tests (i.e., P.835 \cite{reddy2022dnsmos}). Specifically, we use the metric measuring the overall signal quality \textit{OVRL}, and specifically the speech quality \textit{SIG}. Note here that intrusive metrics cannot be used to evaluate in-the-wild recordings, like our dataset, as the reference is not available due to the lack of parallel data. Statistical significance for improved performance is estimated using one-tailed $t$-test on error distributions, asserting significance for \emph{p}-values $\leq0.05$.







\section{Results and Discussion}
\label{sec:results}

\begin{figure}[t!]
\centering
\includegraphics[width=\columnwidth]{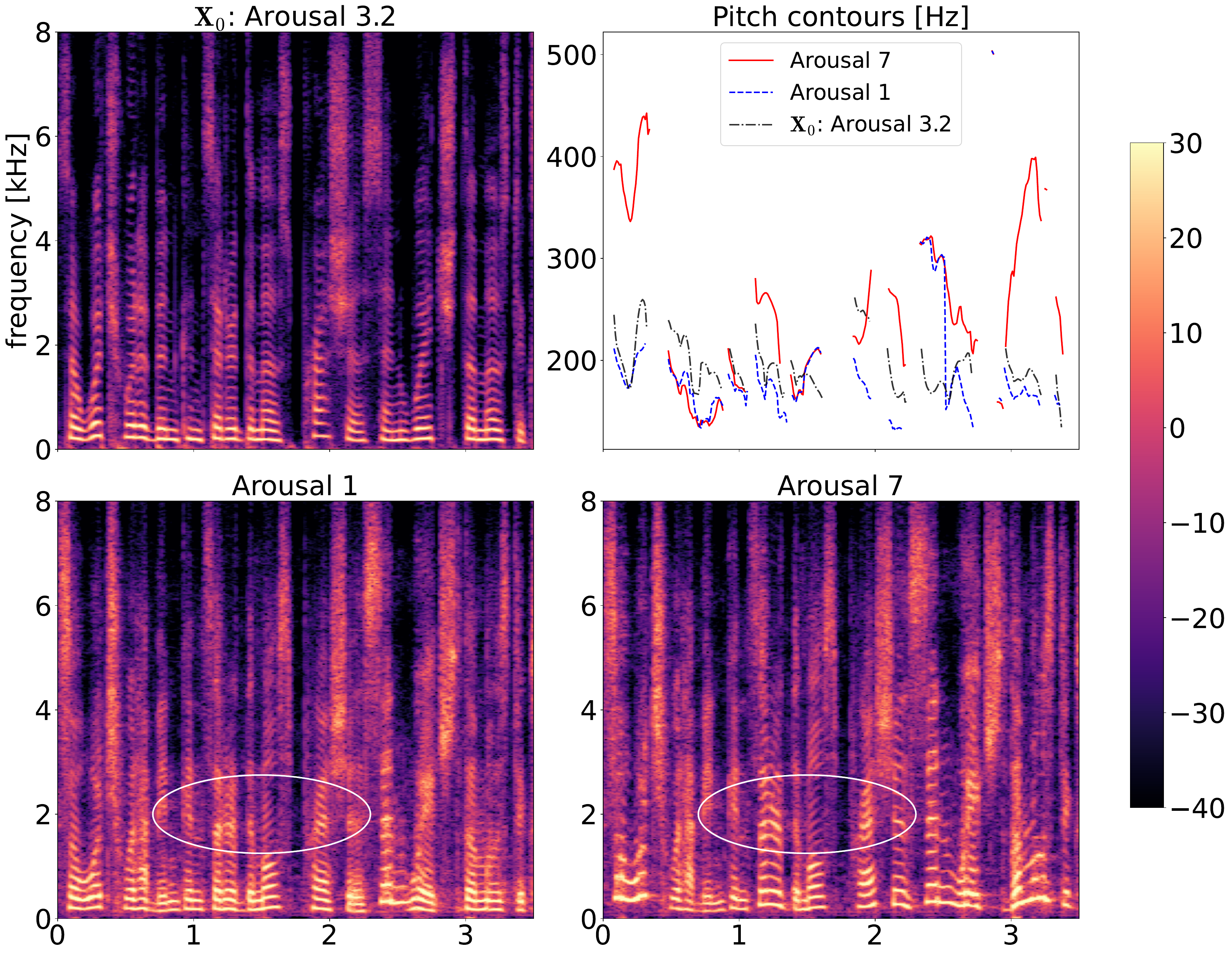}
	\caption{Sample \text{log}-energy spectrogram of emotion converted speech, along with comparisons on pitch contours.}
	\label{fig:sample_spec}%
\end{figure} 

\noindent\textbf{Overall performance:} We validate the overall performance of the proposed \textit{EmoConv-Diff} against a baseline, the HiFiGAN-based SEC system \cite{hifigan_rajprabhu23}, henceforth mentioned as \textit{HiFiGAN}, which to the best of our knowledge is the only prior work on SEC using in-the-wild and non-parallel data. In addition to HiFiGAN \cite{hifigan_rajprabhu23}, we use three different versions of the EmoConv-Diff, (i) $\mathcal{L}_{s}$, which is only trained on the score matching loss $\mathcal{L}_s$, (ii) $\mathcal{L}_{s} + \mathcal{L}_{m}(\bX_t)$, which also uses the mel reconstruction loss $\mathcal{L}_{m}$ tuned on $\bX_t$, and (iii) $\mathcal{L}_{s} + \mathcal{L}_{m}(\bX_0)$, where the mel reconstruction loss $\mathcal{L}_{m}$ is tuned on the approximated source mel spectrogram $\hat{\bX}_0$ \eqref{eq:X0}. 

From the results presented in Table \ref{tab:overall_compare}, we note the following. First, the EmoConv-Diff version $\mathcal{L}_{s} + \mathcal{L}_{m}(\bX_0)$ achieves the best SER errors with statistical significance, achieving $\mathcal{L}_{mse}$ of 0.072 and $\mathcal{L}_{abs}$ of 21$\%$. This confirms the emotion conversion capability of the proposed diffusion model. Second, in terms of the speech quality and overall signal quality, the EmoConv-Diff version $\mathcal{L}_{s} + \mathcal{L}_{m}(\bX_0)$ performs on par with the HiFiGAN baseline. The variant achieves speech quality performance of 3.21 SIG and an overall signal quality of 2.78 OVRL. Third, the introduction of the mel reconstruction loss $\mathcal{L}_{m}(\hat{\bX}_0)$ tuned on the derived approximation of source $\bX_0$ \eqref{eq:X0} improves the performance of the diffusion model, in terms of both the DNSMOS scores and SER errors. Finally, when the mel reconstruction loss $\mathcal{L}_{s}$ is tuned on $\bX_t$, the performance in terms of the SER errors diminishes, signifying the noisy nature of $\bX_t$ and the efficiency of the derived $\hat{\bX}_0$ during the training phase.

\noindent\textbf{Qualitative analysis of spectrograms:} Fig.~\ref{fig:sample_spec} shows sample spectrograms of the source speech $\bX_0$ of arousal $e=3.20$, the converted speech of \textit{reduced} arousal $\Bar{e}=1$, and of \textit{increased} arousal $\Bar{e}=7$. A high average pitch of the speech signal is directly associated with a high intensity of emotion \cite{zhou2023_emointensitycontrol, hifigan_rajprabhu23}. Therefore, in Fig.~\ref{fig:sample_spec}, we also plot the pitch contours of the respective converted speech and the source $\bX_0$. Comparing the spectrograms of arousal $1$ and arousal $7$, from the marked eclipses, we can observe that for increased arousal of $7$ the spectrograms have larger magnitudes in the mid-frequencies. This reveals that the proposed EmoConv-Diff model associates \textit{larger frequency magnitudes} for high arousal speech than for low arousal speech. From the pitch contours, it can be further noted that the synthesized speech for high arousal ($\Bar{e}=7$) has a \textit{higher mean and variability of pitch}, than that of both the ground-truth speech ($e=3.20$) and the synthesized speech for low arousal ($\Bar{e}=1$). This difference in pitch is also clearly notable in the audio examples available online\footnote{\url{https://uhh.de/inf-sp-emoconvdiff} \label{link:audio_sample}}. These results show that the proposed model successfully performs SEC by aptly conditioning on the emotion content.


\begin{figure}[t!]
\centering
    \begin{subfigure}{0.235\textwidth}
        \centering
    \includegraphics[width=\textwidth]{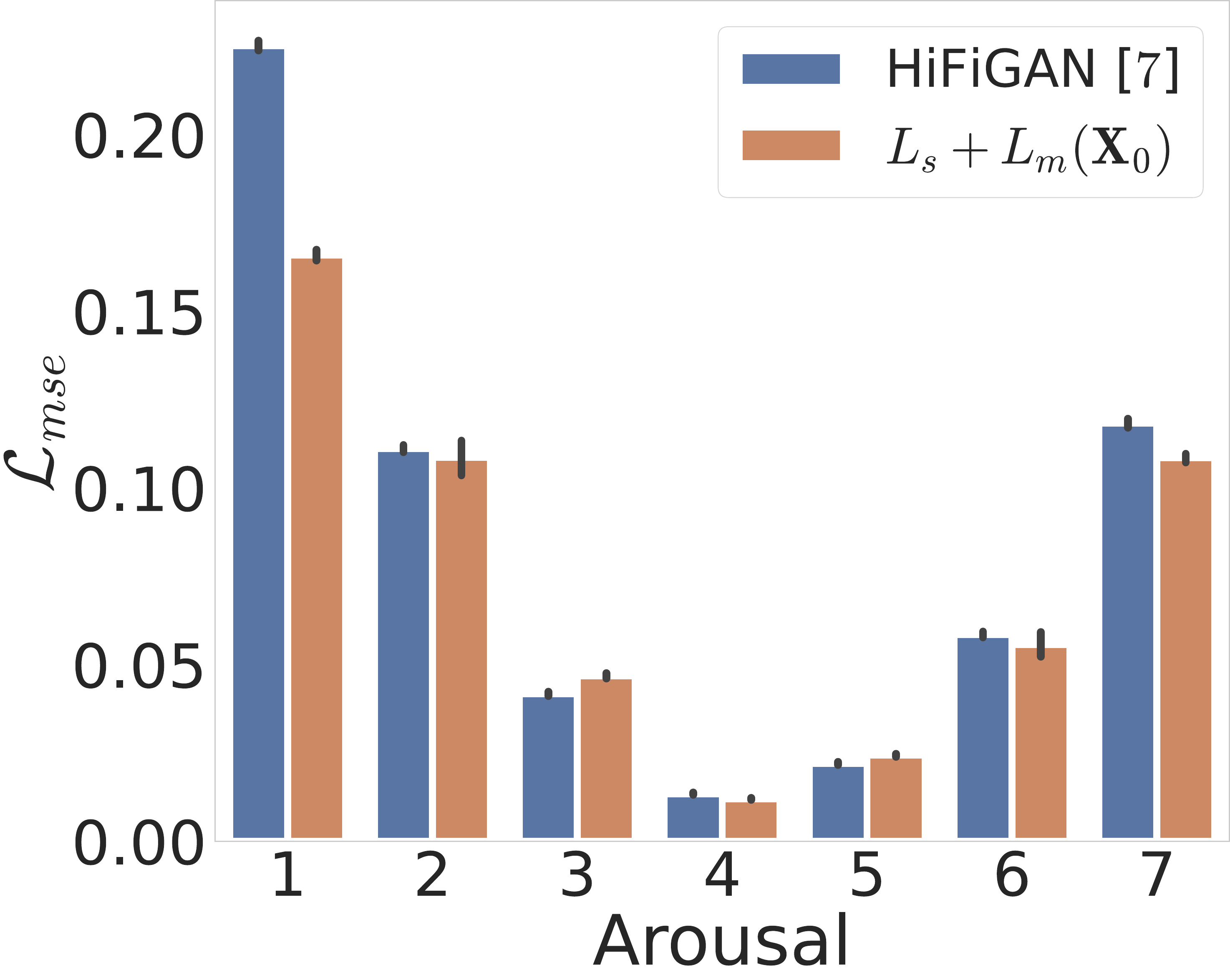}
        \subcaption{Target arousal $\Bar{e}$}
        \label{fig:serloss_trends_target}
    \end{subfigure}
\centering
    \begin{subfigure}{0.235\textwidth}
        \centering
        \includegraphics[width=\textwidth]{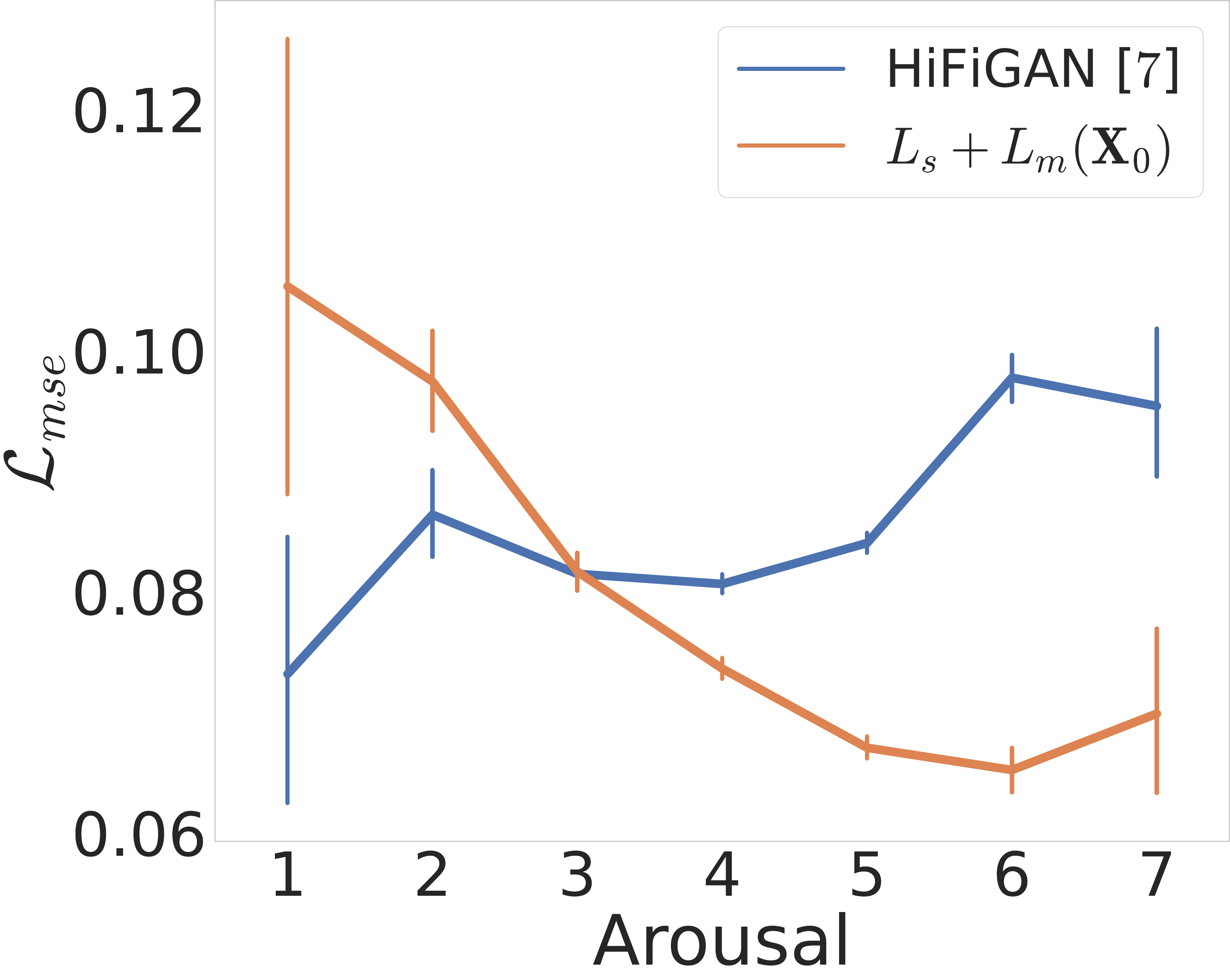}
        \subcaption{Source arousal $e$}
        \label{fig:serloss_trends_source}
    \end{subfigure}
    \caption{Class-wise $L_{mse}$ performances for target arousal $\Bar{e}$ and ground-truth arousal $e$.}
    \label{fig:arousal_classwise_trends}
\end{figure}


\noindent\textbf{Performance for target arousal $\Bar{e}$:} SEC systems generally tend to perform well on certain emotion pairs and emotion classes. For example, \cite{rizos2020stargan} notes that the emotional pairing of "angry" and "sad" is easier to convert than the pairing of "happy" and "angry". Moreover, given that the emotion classes are imbalanced in in-the-wild datasets \cite{lotfian2017building}, with fewer samples along the extremes of emotion scale, SEC for extreme values of $\Bar{e}$ is a general challenge \cite{hifigan_rajprabhu23, wang23ka_interspeech}. To investigate this, in Fig. \ref{fig:serloss_trends_target}, we plot the $\mathcal{L}_{mse}$ performance with respect to each of the target arousal classes $\Bar{e}$. From the plot, we note that the proposed EmoConv-Diff model makes the \textit{largest} improvements along the extreme target arousal values (i.e., $\Bar{e}=$1 and 7) while performing on par along the mid scale arousal values (i.e., $\Bar{e}=$2, 3, 4, 5 and 6). This confirms that the proposed EmoConv-Diff overcomes a crucial shortcoming of existing SEC systems by improving along the extreme values of $\Bar{e}$.



\noindent\textbf{Performance for source arousal ${e}$:} While it is important to evaluate the SEC performance with respect to the target arousal $\Bar{e}$, it is also important to the SEC performance with respect to the arousal of the source speech $\bX_0$ (i.e., $e$). In Fig. \ref{fig:serloss_trends_source}, we also plot the $\mathcal{L}_{mse}$ performances with respect to $e$ and observe the following. First, for both the HiFiGAN \cite{hifigan_rajprabhu23} and the proposed EmoConv-Diff, the standard deviation of $\mathcal{L}_{mse}$ with respect to the extreme emotions ($e=$1 and 7) is \textit{larger} than the mid scale values of $e$. This indicates that it is generally harder for SEC systems to convert the emotion of source $\bX_0$ with already extreme emotions (i.e., $e=$1 and 7), while it is easier to convert the emotion neutral emotion source $\bX_0$ (i.e., $e=$3, 4 and 5). Second, we note contrasting behaviors between the HiFiGAN and the EmoConv-Diff. While the EmoConv-Diff achieves better performance for \textit{higher} source arousal values ($e>$3) than \textit{lower} arousal values, the HiFiGAN does better for \textit{lower} source arousal values ($e<$3) than \textit{higher} arousal values. Moreover, the proposed EmoConv-Diff model performs better than the HiFiGAN in four of the seven arousal classes, which points to the superior SEC capability of the EmoConv-Diff compared to the HiFiGAN baseline.

\section{Conclusion}
\label{sec:conclusion}
Emotion-conditioned speech synthesis (ESS) is an important application that can promote the naturalness of machine communication. Speech emotion conversion (SEC) is a sub-field of ESS. In this paper, we moved beyond the typical reliance on acted-out data sets and parallel samples in SEC, by proposing a diffusion-based generative model and using the continuous arousal dimension to represent emotions while also achieving intensity control. We validated our model using the MSP-Podcast v1.10, a large in-the-wild dataset. We show that our proposed diffusion model, the EmoConv-Diff, is indeed able to synthesize speech for a controllable target emotion. In particular, in comparison to our prior work \cite{hifigan_rajprabhu23}, our model shows improved performance along the extreme values of arousal and thereby addresses a common challenge in the SEC literature \cite{hifigan_rajprabhu23, wang23ka_interspeech}.




\vfill\pagebreak

\bibliographystyle{ieeetr}
\bibliography{refs}

\end{document}